\documentclass[]{spie}  

\usepackage[colorlinks=true, allcolors=blue]{hyperref}

\usepackage{amssymb}
\usepackage{adjustbox}
\usepackage{graphicx}
\usepackage{siunitx}
\usepackage{mathtools}
\usepackage{threeparttable}
\usepackage{float}
\usepackage{booktabs}
\usepackage{tabularx}
\usepackage{makecell}
\newcolumntype{C}[1]{>{\centering\arraybackslash}p{#1}}
\newcolumntype{L}[1]{>{\raggedright\arraybackslash}p{#1}}
\usepackage{multicol}
\usepackage[T1]{fontenc}
\usepackage{csquotes}
\usepackage{url}
\usepackage{subcaption}
\DeclareMathOperator{\E}{\mathbb{E}}

\title{Universal Lymph Node Detection in Multiparametric MRI with Selective Augmentation}

\author[1]{Tejas Sudharshan Mathai}
\author[1]{Sungwon Lee}
\author[1]{Thomas C. Shen}
\author[2]{\\Zhiyong Lu}
\author[1]{Ronald M. Summers}

\affil[1]{Imaging Biomarkers and Computer-Aided Diagnosis Laboratory, Radiology and Imaging Sciences, Clinical Center, National Institutes of Health, Bethesda, MD, USA}
\affil[2]{National Center for Biotechnology Information, National Library of Medicine, National Institutes of Health, Bethesda, MD, USA}

\authorinfo{Send correspondence to T.S.M.: tejas dot mathai at nih dot gov}

\graphicspath{ 
{fig_money/} 
{fig_money/13561_5_24_13561_6_b0_24_labmda_35_ok/}
{fig_results/}
{fig_results/15235_10_17_15235_11_b0_17_labmda_35_ok/}
}

\begin{document} 
\maketitle


\begin{abstract}
Robust localization of lymph nodes (LNs) in multiparametric MRI (mpMRI) is critical for the assessment of lymphadenopathy. Radiologists routinely measure the size of LN to distinguish benign from malignant nodes, which would require subsequent cancer staging. Sizing is a cumbersome task compounded by the diverse appearances of LNs in mpMRI, which renders their measurement difficult. Furthermore, smaller and potentially metastatic LNs could be missed during a busy clinical day. To alleviate these imaging and workflow problems, we propose a pipeline to universally detect both benign and metastatic nodes in the body for their ensuing measurement. The recently proposed VFNet neural network was employed to identify LN in T2 fat suppressed and diffusion weighted imaging (DWI) sequences acquired by various scanners with a variety of exam protocols. We also use a selective augmentation technique known as Intra-Label LISA (ILL) to diversify the input data samples the model sees during training, such that it improves its robustness during the evaluation phase. We achieved a sensitivity of $\sim$83\% with ILL vs. $\sim$80\% without ILL at 4 FP/vol. Compared with current LN detection approaches evaluated on mpMRI, we show a sensitivity improvement of $\sim$9\% at 4 FP/vol. 
\end{abstract}

\keywords{MRI, Multi-Parametric, T2, DWI, Lymph Node, Detection, Selective Augmentation, Deep Learning}

\section{Introduction}
\label{sec_intro}

Lymph nodes (LNs) are small glands that are a part of the lymphatic system and are scattered throughout the body. They contain lymphocytes that travel through the nodal network in search of certain target proteins, which need to be removed from the body. In patients with lymphadenopathy, there is an abnormal proliferation of lymphocytes \cite{Maini2021} that could be caused by many factors, such as infections, autoimmune disease, malignancy among others. For these patients, enlarged and metastatic nodes need to be distinguished from benign nodes \cite{Taupitz2007}. Multi-parametric MRI (mpMRI) is used for LN examination and various sequences are obtained, such as T2 fat suppressed (T2FS) images, diffusion weighted imaging (DWI), and Attenuation Diffusion Coefficient (ADC) maps. AJCC guidelines \cite{Amin2017} provide recommendations on the location and number of LNs to be evaluated for patient treatment. Radiologists routinely measure the size of LNs with the long and short axis diameters (LAD and SAD) to determine malignancy. Nodes with SAD $\geq$ 1cm are considered suspicious for metastasis. Correlation with different series (e.g., T2FS and DWI) is typically sought for malignancy confirmation. 

However, this determination is rendered challenging due to the multitude of imaging scanners, exam protocols in use, observer measurement variability, and institutional guidelines among others. Further complicating the assessment is the diverse appearances and shapes of LNs in mpMRI. Moreover, as sizing nodes is a routine and repetitive task in a radiologist's workflow, some suspicious nodes can be missed during the course of a busy clinical day. To alleviate these imaging and workflow related issues, a number of lymph node (LN) detection algorithms have been published in literature \cite{Zhao2020_mri,Lu2018_mri,Debats2019_mri,Mathai2021_mlmi,Mathai2021_spie,Wang2022_mri,Mathai2022_CARS}. Some of these approaches focus on detecting LNs in specific regions of body (pelvis \cite{Debats2019_mri} and rectal \cite{Zhao2020_mri} areas), while others universally detect both benign and malignant nodes in the body \cite{Lu2018_mri,Mathai2021_mlmi,Mathai2021_spie,Wang2022_mri,Mathai2022_CARS}. However, very few of these approaches \cite{Zhao2020_mri,Lu2018_mri} focus their attention on using mpMRI for LN detection. Difficulty in obtaining retrospective mpMRI studies with different series for algorithmic development is one potential reason. 

\begin{figure}[!t]
\centering
\begin{subfigure}[b]{0.24\columnwidth}
\vspace*{\fill}
  \centering
  \includegraphics[width=\columnwidth,height=2.6cm]{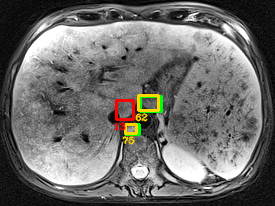}
  \centerline{(a) }
  \includegraphics[width=\columnwidth,height=2.6cm]{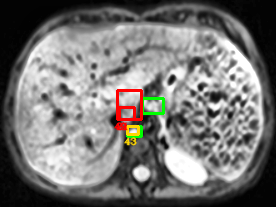}
  \centerline{(e) } 
\end{subfigure} 
\begin{subfigure}[b]{0.24\columnwidth}
\vspace*{\fill}
  \centering
  \includegraphics[width=\columnwidth,height=2.6cm]{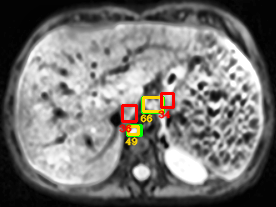}
  \centerline{(b) }
  \includegraphics[width=\columnwidth,height=2.6cm]{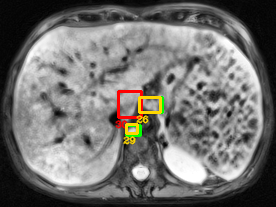}
  \centerline{(f) }
\end{subfigure} 
\begin{subfigure}[b]{0.24\columnwidth}
\vspace*{\fill}
  \centering
  \includegraphics[width=\columnwidth,height=2.6cm]{v_c_21_noBlend}
  \centerline{(c) }
  \includegraphics[width=\columnwidth,height=2.6cm]{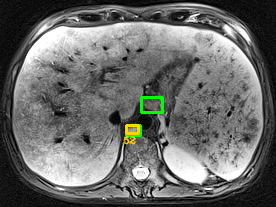}
  \centerline{(g) }
\end{subfigure} 
\begin{subfigure}[b]{0.24\columnwidth}
\vspace*{\fill}
  \centering
  \includegraphics[width=\columnwidth,height=2.6cm]{v_c_21_blend}
  \centerline{(d) }
  \includegraphics[width=\columnwidth,height=2.6cm]{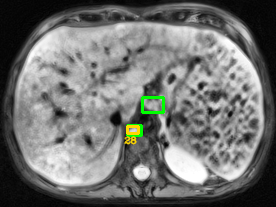}
  \centerline{(h) }
\end{subfigure} 
\caption{LN detection results of a VFNet model are shown in (a)-(f). Green boxes: ground truth, yellow: true positives, and red: false positives. (a) shows a T2FS slice and (b) shows a DWI slice from a mpMRI study. (c) and (d) show results from a 2 T2FS + 1 DWI slice combination; (c) shows results without Intra-Label LISA (ILL) on a T2FS slice, while (d) shows results with ILL on an interpolated slice. Notice the prominent granularity of the spleen visible in (d) due to interpolation of two domains (T2FS and DWI). (e) and (f) display results from a 1 T2FS + 2 DWI slice combination; (e) shows results without ILL on a DWI slice, while (f) shows results with ILL on an interpolated slice. Note that FP were detected in all images except for (d). (g) and (h) show the results of a comparative Faster RCNN baseline \cite{Zhao2020_mri} on a 2 T2FS + 1 DWI slice combination; (g) shows results without ILL on a T2FS slice, while (h) shows results with ILL on an interpolated slice. Note that a LN was missed in (g) and (h) that was captured in (d). VFNet trained with ILL showed higher recall with fewer FP.}
\label{fig:money}
\end{figure}

\begin{figure}[!h]
\centering
\begin{subfigure}[b]{0.24\columnwidth}
\vspace*{\fill}
  \centering
  \includegraphics[width=\columnwidth,height=3cm]{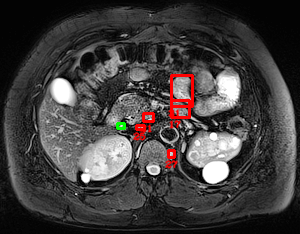}
  \centerline{(a) }
  \includegraphics[width=\columnwidth,height=3cm]{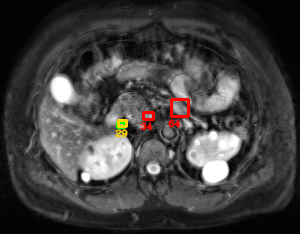}
  \centerline{(b) } 
\end{subfigure}
\begin{subfigure}[b]{0.24\columnwidth}
\vspace*{\fill}
  \centering
  \includegraphics[width=\columnwidth,height=3cm]{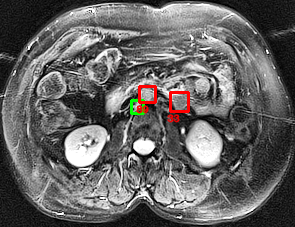}
  \centerline{(c) } 
  \includegraphics[width=\columnwidth,height=3cm]{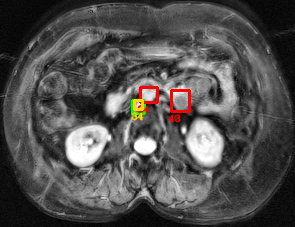}
  \centerline{(d) }
\end{subfigure} 
\begin{subfigure}[b]{0.24\columnwidth}
\vspace*{\fill}
  \centering
  \includegraphics[width=\columnwidth,height=3cm]{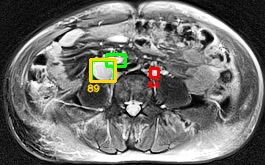}
  \centerline{(e) }
  \includegraphics[width=\columnwidth,height=3cm]{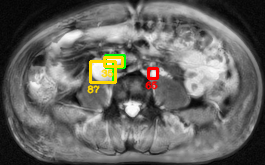}
  \centerline{(f) }
\end{subfigure} 
\begin{subfigure}[b]{0.24\columnwidth}
\vspace*{\fill}
  \centering
  \includegraphics[width=\columnwidth,height=3cm]{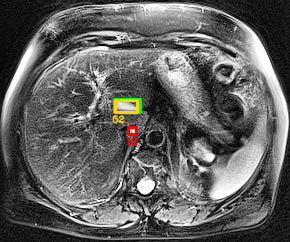}
  \centerline{(g) }
  \includegraphics[width=\columnwidth,height=3cm]{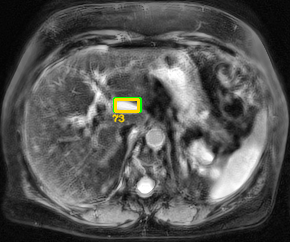}
  \centerline{(h) }
\end{subfigure} 
\caption{Output of the VFNet model on slices from different mpMRI studies. The model was trained with the 2 T2FS and 1 DWI slice combination. The top and bottom rows show outputs of VFNet in $E_{21}$ trained without and with ILL respectively. The top row shows only T2FS slices, while the bottom row shows interpolated slices. Green boxes: ground truth, yellow: true positives, and red: false positives. LN of different sizes (SAD $\geq$ 3mm) that were missed in (a), (c) and (e) were captured in (b), (d) and (f) respectively. (b) and (h) also saw a reduction in the number of FP in contrast to (a) and (g).}
\label{fig:results}
\end{figure}

In this work, we propose an automated CAD pipeline consisting of a VFNet \cite{Zhang2021_vfnet} neural network to universally detect both benign and metastatic nodes in mpMRI studies of the body for their subsequent measurement. Data acquired by various imaging scanners and a variety of exam protocols were used along with a selective augmentation technique known as Intra-Label LISA (ILL) to diversify the input data samples the model sees during training. Contrary to prior work \cite{Zhao2020_mri}, we use full-size inputs for training and testing, and achieve an improvement in sensitivity of $\sim$9\% at 4 FP/vol. We also show a increase in recall of $\sim$83\% with ILL vs. $\sim$80\% without ILL at 4 FP/vol. The use of mpMRI increased recall at 4 FP/vol by $\sim$18\% when compared to only using T2FS as in Wang et al \cite{Wang2022_mri}. As seen in Fig.~\ref{fig:money} and Supplementary Fig.~\ref{fig:results}, our results indicated that a model trained with the 2 T2FS and 1 DWI slice combination and ILL yielded the best LN detection performance.

\section{Methods}
\label{sec_methods}

\textbf{Data.} The Picture Archiving and Communication System (PACS) at our institution was queried for patients who had undergone MRI imaging between January 2015 and September 2019. Originally, a total of 500 multiparametric MRI studies were identified as containing benign and/or malignant nodes and they contained various sequences, such as T2 weighted (T2WI) series, T2 fat suppressed (T2FS) series, diffusion weighted imaging (DWI) and apparent diffusion coefficient (ADC) maps. These studies were acquired using a variety of imaging scanners (GE, Siemens, Philips) and exam protocols. The radiology report associated with a study was also obtained, and a natural language processing algorithm \cite{Peng2020} extracted the LN extent and size measurements. At our institution, LN were measured with either the long axis diameter (LAD) or short axis diameter (SAD), or both simultaneously. A quality check was conducted by a radiologist to standardize the annotations, such that both LAD and SAD were available. Next, studies containing both T2FS and DWI series were identified, and the DWI series was registered to the T2FS series using an ITK-based rigid registration method to have the same origin, resolution, and spacing. This process resulted in 279 studies (n = 279 patients) containing matched and co-registered T2FS and DWI series. The studies had DWI series with multiple b-values. Diffusion effects are more pronounced with high b-values and they result in images with high voxel intensities for LN in contrast to the surrounding (background) tissue. In this work, we exploited all available b-value sequences. These studies were randomly divided into $\sim$69\% train (191 studies, 263 slices, 271 LN), $\sim$8\% validation (22 studies, 27 slices, 29 LN), and $\sim$23\% test (66 studies, 450 slices, 716 LN) splits. The 3D extent of all LN (SAD $\geq$ 3mm) in the test set were fully labeled, while the train and validation splits consisted of only key 2D slices in a 3D volume that were annotated by the original radiologist. Following this division, N4 bias normalization \cite{Tustison2010}, normalization to [1\%, 99\%] of the voxel intensity range \cite{Kociolek2020}, and histogram equalization \cite{Chen2015_histEq} were used to boost the contrast between bright and dark structures in the volumes. The resulting series had various dimensions in the range from (256 $\sim$ 640) $\times$ (192 $\sim$ 640) $\times$ (18 $\sim$ 60) voxels.

\noindent
\textbf{Model.} A recently proposed one-stage object detector called Varifocal Network (VFNet) \cite{Zhang2021_vfnet} was used to detect LN and predict their bounding box coordinates. For more details on the model, we refer the reader to prior work \cite{Zhang2021_vfnet} and to the supplementary material where a brief description of the model architecture and its implementation is provided. After the model had been trained, Weighted Boxes Fusion (WBF) \cite{Solovyev2021} was used to combine the abundant predictions from multiple epochs of a single model run or from multiple runs of a model. 

\noindent
\textbf{Selective Augmentation.} We use a recently proposed method called LISA \cite{Yao2022_LISA} to learn invariant predictors using a selective augmentation approach that is rooted in the MixUp \cite{Zhang2018_mixup} technique. MixUp linearly interpolates training samples in order to remove spurious correlations \cite{Cramer2016} between the domain and labels. Specifically in this work, we use Intra-Label LISA (ILL) to interpolate training samples that have the same label but are sampled from different domains (T2 MRI and DWI). Formally, assume that two data samples (${x}_{i}$, ${y}_{i}$, ${d}_{i}$) and (${x}_{j}$, ${y}_{j}$, ${d}_{j}$) are drawn from two distinct domains ${d}_{i}$ and ${d}_{j}$. Two samples can be linearly interpolated according to:

\begin{equation} 
\label{eq:mixup}
{x}_{m} = \lambda{x}_{i} + (1-\lambda){x}_{j} \quad\mathrm{and}\quad 
{y}_{m} = \lambda{y}_{i} + (1-\lambda){y}_{j}
\end{equation}
\begin{equation}
\label{eq:ERM}
\hat{\theta} := \underset{\theta \in \Theta}{\mathrm{argmin}}\, \E_{\{ (x_{i},y_{i},d_{i}), (x_{j},y_{j},d_{j}) \sim \hat{P} \}} \big[l(f_{\theta}(x_{m}), y_{m})\big]
\end{equation}

\noindent
where $\lambda \in [0,1]$ is the interpolation ratio sampled from a Beta distribution $Beta(\alpha,\beta)$. As this formulation was originally utilized for classification, we re-purpose it for detection in which the label is the same ${y}_{i} = {y}_{j}$. In this work, as mpMRI sequences are co-registered (see Sec.~\ref{sec_expResults}), the label is a LN bounding box. This results in interpolated samples in which both domains are partially present and any spurious correlations that exist between the domains and labels are removed. Once the inputs are interpolated, an empirical risk minimization setting arises as in Eqn.~\ref{eq:ERM} where given a training distribution ${P}_{tr}$, a loss function $l$ is used to train a model ${f}_{\theta}$ to optimize its parameters $\theta \in \Theta$. Through this process, the model sees diverse training examples and the robustness to noise is improved during the test time evaluation.

\section{Experiments and Results}
\label{sec_expResults}

\noindent
\textbf{Experiments.} Radiologists size LN in studies by scrolling back and forth across the slices in a volume and make annotations on a single key slice. From prior work \cite{Debats2019_mri}, the in-plane slice provided salient information and we mimicked their approach by using a 2.5D image containing three consecutive mpMRI slices with the key slice in the middle for training the VFNet model. However, in order to compare our results against prior work \cite{Zhao2020_mri}, we constructed four experiments with four distinct combinations of T2FS and DWI slices including: 1) 3-slices of only T2FS (${E}_{T}$), 2) 3-slices of only DWI (${E}_{D}$), 3) 1-slice of T2FS and 2-slices of DWI (${E}_{12}$), and 4) 2-slices of T2FS and 1-slice of DWI (${E}_{21}$). Additionally, we carried out another experiment in which we compared the effects of using ILL specifically for the last two combination modes (${E}_{12}$ and ${E}_{21}$). 

\noindent
\textbf{Baseline comparison.} While the slices in a comparative baseline \cite{Zhao2020_mri} were cropped to 256$\times$256 pixels encompassing LN in the rectal region, we did not crop our slices and used the full-sized images as training inputs. They also used a Mask RCNN model for detection and segmentation, but we did not have segmentation labels in this work. To perform a fair comparison, we re-implemented their work with the same hyper-parameters, but without cropping, using a Faster RCNN \cite{Ren2015_fasterrcnn} model.

\noindent
\textbf{Results.} Similar to prior work \cite{Zhao2020_mri,Wang2022_mri}, a clinically acceptable result for LN detection meant a sensitivity of 65\% at 4-6 FP per volume. From Table~\ref{table_LN_detection_results}, we can see that the mAP and sensitivities are significantly higher for the experiment ${E}_{T}$ (only T2FS slices) in contrast to the experiment ${E}_{D}$ (only DWI slices). We believe that the diffuse appearance of tissue structures in DWI sequences was detrimental to LN localization. In our experiment ${E}_{12}$ (1 T2FS and 2 DWI slice combination), the VFNet model performed worse when compared with the model from experiment ${E}_{21}$ (2 T2FS and 1 DWI slice combination). Recalls were lower for ${E}_{12}$ although the mAP was similar ($3^{rd}$ vs. $5^{th}$ row). In experiment ${E}_{21}$, VFNet trained with ILL showed improvements compared to a model trained without ILL ($5^{th}$ vs. $6^{th}$ row). The same trend held for ${E}_{12}$, although the sensitivity marginally improved and the mAP was almost similar. This strengthened our belief that a reliance on the DWI series did not improve LN detection due to the diffuse nature of tissue structures. Moreover, ${E}_{21}$ showed significant performance gains when compared with ${E}_{T}$ and ${E}_{D}$ respectively. These results indicated the complementary nature of DWI and T2FS when T2FS is predominantly represented in the input, and the robustness of our VFNet model at test time through selective augmentation with ILL. 

\begin{table}[!t]
\centering\fontsize{9}{12}\selectfont 
\setlength\aboverulesep{0pt}\setlength\belowrulesep{0pt} 
\setlength{\tabcolsep}{7pt} 
\setcellgapes{3pt}\makegapedcells 
\caption{Performance comparison of our VFNet detector against other methods. ``Exp'' stands for an experiment with one of four domain \{T2FS, DWI\} combination modes. ``S'' stands for Sensitivity @[0.5, 1, 2, 4, 6, 8] FP. ``NSA'' indicates no selective augmentation. ``ILL'' stands for Intra-Label LISA. ``--'' is unavailable.}
\begin{adjustbox}{max width=\textwidth}
\begin{tabular}{@{} c|c|c|c|c|c|c|c|c|c|c @{}} 
\toprule
\#      &   Method          & Exp                             & Mode                          & mAP       & S@0.5     & S@1       & S@2       & S@4       & S@6       & S@8       \\
\midrule

1       &   VFNet          & $E_{T}$                           & T2FS Only                     & 51.7      & 47.2      & 57.1      & 71        & 80.7      & 82.4      & 85.3      \\
2       &   VFNet          & $E_{D}$                           & DWI Only                      & 39.2      & 34.9      & 45.5      & 59.4      & 71        & 77.5      & 79.3      \\

3       &   VFNet          & $E_{12}$                           & NSA            & 53.7      & 48        & 58.8      & 70.4      & 79.2      & 83.9      & 86.9      \\
4       &   VFNet          & $E_{12}$                           & ILL          & 53.3      & 48.2      & 59.8      & 71.5      & 80.4      & 84.5      & 87.4      \\
5       &   VFNet          & $E_{21}$                           & NSA            & 53.8      & 47.6      & 60.8      & 71.6      & 79.9      & 83.8      & 85.8      \\
6       &   VFNet          & $E_{21}$                           & ILL          & \textbf{55.8}      & \textbf{50.3}      & \textbf{62.7}      & \textbf{72.5}      & \textbf{82.4}      & \textbf{86.6}      & \textbf{89.2}      \\

7       &   Faster RCNN          & $E_{12}$                     & NSA            & 31.4      & 29.5      & 41.1      & 52.9      & 67.6      & 72.9      & 76.4      \\
8       &   Faster RCNN          & $E_{12}$                     & ILL          & 34.7      & 32.7      & 44.8      & 56        & 71.8      & 76.3      & 79.7      \\
9       &   Faster RCNN          & $E_{21}$                     & NSA            & 37.1      & 34.4      & 43        & 55.9      & 67.7      & 72.1      & 76.3      \\
10       &   Faster RCNN          & $E_{21}$                     & ILL          & \textbf{40.8}      & \textbf{35.6}      & \textbf{45.9}      & \textbf{57.7}      & \textbf{73}        & \textbf{78.1}      & \textbf{80.4}      \\

\midrule

11       &   Zhao 2020 \cite{Zhao2020_mri} (3D)          & $E_{12}$      & NSA        & 73.5      & --        & --        & --        & --        & --        & 80   \\
12       &   Zhao 2020 \cite{Zhao2020_mri} (3D)          & $E_{21}$      & NSA        & 59.7      & --        & --        & --        & --        & --        & 81.3   \\
13       &   Wang 2022 \cite{Wang2022_mri} (3D)          & $E_{T}$       & T2FS Only                 & --        & --        & --        & --        & 64.6      & --        & --    \\
14       &   Mathai 2022 \cite{Mathai2022_CARS} (3D)     & $E_{T}$       & T2FS Only                 & 52.3      & 46.5      & 58        & 68.9      & 78.7      & 82.7      & 85.2    \\

\bottomrule
\end{tabular}
\end{adjustbox}
\label{table_LN_detection_results}
\end{table}

We also compared our results against those from prior work \cite{Zhao2020_mri,Wang2022_mri,Mathai2022_CARS}. In previous LN detection work \cite{Zhao2020_mri}, experiments ${E}_{12}$ and ${E}_{21}$ were conducted on data acquired from patients with rectal adenocarcinoma. Their results ($11^{th}$ and $12^{th}$ rows) were obtained on mpMRI series that were cropped to the rectal region. As these results are not representative, we trained a Faster RCNN model with their provided hyperparameters on our full-size input data. As seen in rows 7 through 10 in Table~\ref{table_LN_detection_results}, ${E}_{21}$ without ILL outperformed ${E}_{12}$ without ILL in both mAP and recalls across the board. Using selective augmentation through ILL to train Faster RCNN yielded improvements in LN detection for both ${E}_{12}$ and ${E}_{21}$ with the best performance obtained in ${E}_{21}$ with ILL. However, the results were still significantly lower than those obtained with VFNet ($6^{th}$ vs. $10^{th}$ row); VFNet sensitivity for ${E}_{21}$ improved by $\sim$9\% at 4 FP/vol and by $\sim$9\% at 8 FP/vol respectively. 

As prior LN detection approaches \cite{Wang2022_mri,Mathai2022_CARS} used only T2FS volumes in their experiments, we first compared our results from ${E}_{T}$ with them. Compared against Wang et al. \cite{Wang2022_mri} ($1^{st}$ vs. $13^{th}$ row), our recall at 4 FP/vol was $\sim$16\% higher. Compared with Mathai et al. \cite{Mathai2022_CARS} ($1^{st}$ vs. $14^{th}$ row), our mAP was slightly lower but recall at 4 FP/vol improved by 2\%. Next, we compared their results against ${E}_{21}$ with VFNet. When compared with Wang et al. \cite{Wang2022_mri} ($6^{th}$ vs. $13^{th}$ row), our recall at 4 FP/vol was $\sim$18\% higher. In contrast to Mathai et al. \cite{Mathai2022_CARS}, our mAP and recall at 4 FP/vol increased by 3.5\% and 3.7\% respectively. This meant that we saw an improvement in LN detection by using a 2 T2FS and 1 DWI slice combination and ILL for selective augmentation. The runtime of our model on a volume was $\sim$2.9 seconds on average.

\section{Discussion and Conclusion}
\label{sec_conclusion}

\textbf{Discussion.} Universal localization of benign and metastatic LNs in the body is critical, as ensuing measurements differentiate metastatic from benign nodes. However, localization and measurement is a repetitive task that is routinely performed by a radiologist in mpMRI studies. It can be sped up through the proposed automated pipeline with a VFNet model that can detect LN with SAD $\geq$ 3mm and runs in $<3$ seconds per volume. In this work, we have seen that a model trained on 2.5D images compiled from a T2FS series fared better in contrast to one trained on images from a DWI series. These results are corroborated in Zhao et al. \cite{Zhao2020_mri}, however their best results were achieved with 1 T2FS and 2 DWI slice combination. They arrived at this conclusion after cropping their input images to the rectal region. But in contrast to their findings, we have identified that a full-sized 2.5D input comprising of 2 T2FS and 1 DWI slice combination worked better for universal LN detection. In our experiments, we also observed ILL improving detection by yielding a mAP of $\sim$56\% and recalls of 82.4\% and 89.2\% at 4 and 8 FP/vol respectively. 

Furthermore, the improvements were higher for the Faster RCNN model trained with ILL for both $E_{12}$ and $E_{21}$ experiments. ILL provided the most benefit for Faster RCNN over VFNet, and the addition of other mpMRI sequences, such as ADC maps, could further enhance the already significant representation capacity of VFNet for LN detection \cite{Mathai2022_CARS}. Moreover in Zhao et al.\cite{Zhao2020_mri}, the data was acquired using only a GE imaging scanner in the rectal region and the described results pertain to only that scanner and anatomical area. However in our work, the data was acquired at the abdomen level (chest, abdomen, pelvis) with a variety of imaging scanners and exam protocols allowing our model to see diverse examples during training, and rendering our results to be more descriptive of real-world performance. Future work is directed towards utilizing the trained model to mine additional LNs in the studies, such that LN detection is improved further.

\medskip
\noindent
\textbf{Conclusion.} In this work, we have described an automated pipeline that consists of a VFNet neural network to detect LNs in mpMRI sequences. The goal of this pipeline is to aid a radiologist in quickly ascertaining the location of LN, such that they can be sized and assessed for lymphadenopathy. Our model is trained on T2FS and DWI data acquired by various scanners and differing exam protocols, and uses a selective augmentation method known as Intra-Label LISA (ILL) to improve the diversity of samples seen during model training. We achieved the best results after training a VFNet model with 2.5D images comprising of a data combination of 2 T2FS slices and 1 DWI slice. Our mAP was $\sim$56\% and recalls at 4 FP/vol and 8 FP/vol were 82.4\% and 89.2\% respectively. With mpMRI series, our recall at 4 FP/vol improved by $\sim$9\% in contrast to Zhao et al \cite{Zhao2020_mri}. We also show that using mpMRI increased recall at 4 FP/vol by $\sim$18\% when compared with Wang et al. \cite{Wang2022_mri}, and in contrast to Mathai et al. \cite{Mathai2022_CARS}, mAP and recall at 4 FP/vol increased by 3.5\% and 3.7\% respectively. Our results showed that a model trained with 2 T2FS and 1 DWI slice combination and ILL yielded the best detection performance.

\acknowledgments 
 
This work was supported by the Intramural Research Programs of the NIH National Library of Medicine and NIH Clinical Center (project number 1Z01 CL040004). 

\bibliography{main_ref.bib} 
\bibliographystyle{spiebib} 

\newpage
\section{SUPPLEMENTARY MATERIAL}
\label{sec:supplementary}

\textbf{VFNet Neural Network Model.} The Varifocal Network (VFNet) \cite{Zhang2021_vfnet} merged a Fully Convolutional One-Stage Object (FCOS) detector \cite{Tian2019_fcos} (without the centerness branch) and an Adaptive Training Sample Selection (ATSS) mechanism \cite{Zhang2020_ATSS}. The model replaced the class label for a predicted bounding box with an intersection-over-union (IoU)-aware classification score (IACS) that merged an object's overlap with its location. A varifocal loss was used to predict the IACS, up-weighting the contribution of positive object candidates and down-weighting negative candidates. Moreover, the output bounding boxes were represented using a 9-coordinate star-shaped representation that reduced the misalignment between the ground truth and the predicted box coordinates. 

\medskip
\noindent
\textbf{Implementation.} 2.5D (3-channel) images were used to train detectors, which were implemented with the mmDetection framework \cite{Chen2019_mmdet}. Outside of ILL, data augmentation was performed: random flips, crops, shifts and rotations in the range of [0, 32] pixels and [0, 10] degrees respectively, random contrast and gamma adjustments. ResNet-50 was the backbone (pre-trained with MS COCO weights) for VFNet, while Faster RCNN used ResNet-101 consistent with the implementation in Zhao et al \cite{Zhao2020_mri}. A grid search was run across the batch size and learning rate parameters to obtain the optimal values; for VFNet, batch size and learning rate was set to 8 and 1e-3 respectively, while for Faster RCNN, it was 4 and 1e-6 respectively. The total training epochs was set to 12. Each model was executed 5 times, and the top-3 checkpoints with the lowest validation loss from each run were chosen for testing. Results presented in Table~\ref{table_LN_detection_results} were an average of 5-fold cross-validation. All experiments were run on a NVIDIA DGX workstation running Ubuntu 18.04LTS with 4 Tesla V100 GPUs. Evaluation was always performed at an IoU threshold of 25\% to be consistent with prior work \cite{Zhao2020_mri,Wang2022_mri,Mathai2022_CARS}.

\end{document}